# Electronic Behavior of Superconducting SmFeAsO$_{0.75}$

Y. Sun[1], Y. Ding[1], B. C. Zheng[1], Z. X. Shi[1*], Z. A. Ren[2]

[*]Corresponding author: Z. X. Shi

1. Department of Physics, Southeast University, Nanjing, China 211189

2. Institute of Physics and Beijing National Laboratory for Condensed Matter Physics, Chinese Academy of Sciences, Beijing 100190, P. R. China

Email: zxshi@seu.edu.cn

**Abstract**

High-quality polycrystalline SmFeAsO$_{0.75}$ was synthesized with a superconducting transition width less than 1 K, and the electronic behavior was systematically studied by transport and specific heat measurements. An obvious superconducting jump was witnessed, together with a very small normalized superconducting jump, $\Delta C/\gamma_n T_c \sim 0.2$, which is much smaller than expected by the BCS theory. A strong temperature dependent Hall coefficient was found and attributed to the partial gapping of the Fermi surface up to the temperature of 160 K which was predicted and supported by the emergence of the pseudogap. The charge-carrier density as well as the effective mass were also obtained and discussed in detail.

PACS numbers: 74.25.Bt, 74.25.Fy, 74.25.Jb, 74.70.Dd

## 1. Introduction

The recent discovery of superconductivity at 26 K in the iron oxypnictide LaFeAs(O, F) [1] has stimulated great interest among the condensed-matter physics community. Tremendous work was carried out, leading to the emergence of novel iron-based superconductor families with different crystal structures: 1111 (REFeAs(O, F)), 122 ((Ba, K)Fe$_2$As$_2$) [2], 111 (LiFeAs) [3] and 11 (Fe(Se, Te)) [4]. Of all the iron oxypnictide superconductors, SmFeAs(OF) has the highest the critical temperature $T_c = 55K$, and a very high upper critical field of 247 T [5], which is encouraging for applications. Point-contact spectroscopy [6], lower critical field measurements [7], scanning tunneling spectroscopy [8] have revealed an unconventional pairing symmetry. Much research is yet needed to clarify the electronic behavior of this material with regards to both applications and the underlying physical mechanism. However, given the lack of a single crystalline sample of sufficient size, and the inhomogeneity of polycrystalline samples, only a few work have been focused on the specific heat or the Hall effect of the SmFeAs(OF) superconductor. Previous specific heat experiments on the F-doped SmFeAsO polycrystalline sample [9] [10] only show an very broad anomaly associated with the superconducting transition due to the inhomogeneity of the superconducting phase. Hall effect measurements on both F-doped [11] and oxygen-deficient [12] polycrystalline SmFeAsO showed a temperature dependent Hall coefficient, while the mechanism was not fully understood, especially with regards to the heavily doped sample where the spin density wave (SDW) was totally suppressed by doping. The dearth of specific heat data on oxygen-deficient REFeAsO further obscures the underlying physics of this novel superconductor. In this work, high-quality polycrystalline $SmFeAsO_{0.75}$ superconductor with a SC transition width less than 1K was synthesized to enable the study of the specific heat and the Hall effect.

## 2. Experimental

Superconducting $SmFeAsO_{0.75}$ was prepared via high-pressure (HP) synthesis. The starting materials were mixed according to the nominal stoichiometric ratio, then thoroughly ground and pressed into pellets. The pellets were sealed in boron nitride crucibles and sintered in the high pressure synthesis apparatus under 6 Gpa and at the temperature of 1250℃ for 2 hours. Details can be found in [13]. Longitudinal and transverse (Hall) resistivities were measured with a Quantum Design physical property measurement system (PPMS) with temperatures down to 2 K and magnetic fields up to 9 T. Specific heat data was also obtained under the applied fields of 0 and 9 T using the PPMS via the relaxation method.

## 3. Results and Discussions

*3.1 Resistive transition and upper critical field*

Figure 1 shows the temperature dependence of the electrical resistivity for $SmFeAsO_{0.75}$ under zero field. The sample has a very sharp SC transition with an onset temperature of ~ 55.2 K which can be clearly seen in the inset of Figure 1. The transition width $\Delta T_c = T(90\%\rho_n) - T(10\%\rho_n)$ is ~ 0.93 K, which is the smallest ever reported for a polycrystalline sample, and near to that of single crystalline $SmFeAsO_{0.85}$ [14]. The very sharp transition indicates a very homogeneous SC phase, although resistivity measurements do not probe the bulk of the sample, just SC percolative paths. The residual resistivity ratio (RRR) = ρ (300K) / ρ ($T_{c,\text{onset}}$) is equal to 5.9, which is even larger than the value of ～ 5 previously seen for a single crystalline sample [14], and this high value further points to the high quality of the sample. The resistivity follows a *T*-linear behavior from $T_c$ to ~ 150 K, and then deviates from this relationship, which is similar to behavior seen in F doped SmFeAsO sample [11], and may originate from the scattering of charge carriers by fluctuation associated with the quantum critical point [15].

The inset of Figure 1 shows the temperature dependence of resistivity with applied fields up to 9 T. As is observed from the figure, the slight downward shift in the onset transition temperature with increasing *H* is accompanied by a much greater downward shift in the zero resistance temperature. This is understandable since the latter is determined by a multitude of factors, including but not limited to anisotropic superconductivity, vortex motion, and weak links between the grains, while the former is mainly associated with $H_{c2}^{ab}$ since grains with their a-b planes parallel to the applied field become superconducting first upon cooling. Taking a criterion of 90% $\rho_n$, the value of the zero-temperature upper critical field $H_{c2}^{ab}(0)$ can be obtained by the Werthamer–Helfand–Hohenberg (WHH) formula $H_{c2}(0) = -0.693 T_c \left.\frac{dH_{c2}}{dT}\right|_{T_c}$.

Taking $T_c$=55.2K, $\left.\frac{dH_{c2}^{ab}}{dT}\right|_{T_c} = -5.92$, the $H_{c2}(0)$ is calculated to be roughly 225.5 T which clearly exceeds the Pauli limit of 145 ~ T obtained from the single crystal data [16], indicating an unconventional superconductivity in the iron-based superconductor [17]. The upper critical field $H_{c2}^{ab} \sim \phi_0 / 2\pi \xi_a \xi_c = \phi_0 / 2\pi \xi_a^2 \Gamma^{-1/2}$, where $\Gamma = m_c/m_{ab} = \rho_c/\rho_{ab} = \xi_{ab}^2/\xi_c^2$ is the effective mass or resistivity anisotropy parameter. Since $\xi_a$ is proportional to $1/T_c$, the slope of $H_{c2}^{ab}$ versus $T$, $\frac{dH_{c2}^{ab}}{dT}$ near $T_c$ is proportional to $T_c \Gamma^{1/2}$. By choosing $\left.\frac{dH_{c2}^{ab}}{dT}\right|_{T_c}$ = -5.92 T/K, $T_c$ = 55.2K from the resistivity transition of our sample, and $\Gamma$ =15 [18], $\left.\frac{dH_{c2}^{ab}}{dT}\right|_{T_c}$ = -2.7 T/K, $T_c$=28 K, for LaFeAs(O, F) [12], together with the relation $\frac{dH_{c2}^{ab}/dT|_{T_c}(Sm)}{dH_{c2}^{ab}/dT|_{T_c}(La)} = \frac{T_c(Sm)\Gamma^{1/2}(Sm)}{T_c(La)\Gamma^{1/2}(La)}$, the effective mass anisotropy parameter $\Gamma$ is estimated to be about 18.6. The value of $\Gamma$ we get is similar to the single crystal results in which $\Gamma$ is between 9 and 36 [16]. The not strong anisotropy, which is even smaller than YBa$_2$Cu$_3$O$_7$ single crystals [19], potentially allows this superconductor to be used for a variety of future applications. The effective mass anisotropy is close to the single crystal NdFeAsO$_{0.82}$F$_{0.18}$, which varies from 19-24, and is also much smaller than the results of the band calculation [17].

*3.2. Specific heat*

Figure 2 shows the 2-160 K specific heat of the SmFeAsO$_{0.75}$ at zero field. This temperature range was chosen to include the structure and SDW transition [9]. No anomaly associated with these effects was observed, implying that the transitions were totally suppressed by oxygen deficiencies, enabling the lattice specific heat to be more reliable. An anomaly associated with the superconducting transition is observed at ~ 55.2 K, in accordance with the resistivity measurement. The superconducting transition can be seen more clearly in the lower inset of Figure 2, which is plotted as *C/T* vs *T*. Such an anomaly was absent or not obvious in previously reported data for LaO$_{1-x}$F$_x$FeAs [20], SmO$_{1-x}$F$_x$FeAs [9] [10] superconducting samples. This also manifests the high quality as well as the homogeneity of the sample, which make the results obtained from the specific heat more reliable. At low temperatures, the specific heat is seen to have a sharp peak at 3.3 K, previously witnessed in the F doped SmFeAsO sample [9] [10] [21], which is replotted in the upper inset of Figure 2. The peak, related to the magnetic ordering of $Sm^{3+}$, changes little under the magnetic field of 9 T which is discussed in detail by Riggs [22].

Due to the specific heat peak at low temperature it's hardly to extrapolate the electronic specific heat coefficient $\gamma$ in the low temperature limit. The low temperature specific heat is replotted as $C/T$ vs $T^2$ in Figure 3. Considering the peak was at a relatively low temperature position and the good linearly behavior of the data between 14 and 28 K, the specific heat can be fitted by $C/T = \gamma + \beta T^2$, which allows one to estimate the value of Sommerfeld coefficient $\gamma$ related to the electronic contribution, and the prefactor β which characterizes the lattice contribution to the specific heat in a simple Debye approximation. The fitting gives $\gamma = 28.12 mJ/molK^2$, $\beta = 0.29 mJ/molK^4$ for $H = 0$ T, and $\gamma = 29.76 mJ/molK^2$, $\beta = 0.29 mJ/molK^4$ for $H = 9$ T. The obtained Sommerfeld coefficient $\gamma$ is usually taken as the residual electronic specific heat which may be the contribution of the electronic excitation from nodal gap superconductor, the quantum fluctuation, as well as the non-superconducting phase. And in our sample the obtained $\gamma$ may also be affected by the peak at low temperature. Compared to the small value of $\gamma = 0.69 mJ/molK^2$ seen in $LaFeAsO_{0.9}F_{0.1}$ [20], the relatively large $\gamma$ of our sample is more likely caused by the combined influence of a non-superconducting phase and the low temperature peak rather than the nodal gap or possible quantum fluctuations. The presence of a non-superconducting phase would only be seen in the calorimetric data which measure the heat capacity of the entirety of the sample; measuring the resistivity transition only gives the transition across a SC path in the sample. Though the value of $\gamma$ may be affected by the abovementioned factors, the value of $\beta$, which largely describes the phononic heat capacity, is relatively constant. Using the relation: $\Theta_D = (\frac{12\pi^4 N_A Z k_B}{5\beta_n})^{1/3}$ where $N_A$ is the Avogadro constant, Z is the number of atoms per formula unit, and $k_B$ is the Boltzmann's constant, we obtain the Debye temperature $\Theta_D = 293 K$ for both $H=0$ and 9 T, which is reasonable, considering that the phonon contribution to the specific heat does not depend on the applied magnetic field. Based on the obtained Debye temperature $\Theta_D$, normal state specific heat can be fitted by the combined Einstein-Debye model [10] [21] [23]: $C(T) = \gamma_n T + A_D C_D(T, \Theta_D) + A_E C_E(T, \Theta_E)$, where $\gamma_n$ is the normal state Sommerfeld coefficient, $C_D$ and $C_E$ are Debye and Einstein function, and $A_D$ and $A_E$ are the prefactor of the Debye and Einstein function respectively. The fitting result is shown as the solid line in Figure 2, which gives $\Theta_E = 207.9 K$, $\gamma_n = 40.08 mJ/molK^2$ similar to previously reported F doped SmFeAsO, $\gamma_n = 42 mJ/molK^2$ [10].

To investigate the superconducting transition, specific heat under the applied field of 9 T is shown in Figure 4 together with the data obtained at 0 T. Although 9 T is far from the upper critical field $H_{c2}$ mentioned previously, the jump associated with the superconducting transition is suppressed and moved to a lower temperature. By subtracting the specific heat C (9T), we plot the specific heat jump as $[C(0T)-C(9T)]/T$ in the inset of Figure 4. This jump represents the SC contribution to the electronic specific heat and begins at about 55.2 K, in accordance with the resistivity measurement, with a width of about 5 K, which is the sharpest has ever been reported in the iron-based "1111" system [9] [20] [10]. The sharpest superconducting transition also manifests the good quality of our sample. Using the maximum value of $[C(0T)-C(9T)]/T = 8.11 mJ/molK^2$ together with the Sommerfeld coefficient $\gamma_n = 40.08 mJ/molK^2$, the value of the normalized jump in the specific heat, $\Delta C/\gamma_n T_c$, is about 0.2, which is much smaller than the BCS value of 1.43. The small value is attested to almost all of the reported iron-based "1111" compounds (LaO$_{1-x}$F$_x$FeAs [20], SmO$_{1-x}$F$_x$FeAs [10], LaFePO$_{1-x}$F$_x$ [24] [25]), but absent in other compounds (iron-based "122" [26] and nickel-based "1111" system [27]). This depression in the magnitude of the SC anomaly is likely to be the result of: (1) Although the resistivity measurement gave out a very sharp transition width less than 1 K as previously mentioned, the specific heat shows a larger transition width more than 1.9 K calculated from the onset and peak temperature from the inset of Figure 4. The broadened transition width will lead to a small obtained specific heat jump $\Delta C$, which will in turn reduce the quantity of $\Delta C/\gamma_n T_c$. (2) The magnetization of the impurities may play the role as a cooper pair breakers which reduce the density of superconducting electrons [24-25] leading to the small normalized specific heat jump. (3) The electronic phase separation as witnessed in the iron-based superconductor [28] [29] may be another possible reason. If the phase separation really exists in the iron-based "1111" system, the percentage of superconducting phase is only about 15% estimated from the maximum doping level of the F or oxygen deficiency [30]. Taking this into consideration, the normalized jump can be corrected as 1.33 which is similar to the BCS value. (4) Lastly, the most likely reason is the emergence of the pseudogap [31] which causes some electrons to be pre-paired before $T_c$, and therefore reduces subsequently $\Delta C$. Furthermore, the pseudogap is also predicted by the Hall measurement which will be discussed later. To confirm the origin of the small normalized specific heat jump, more research is necessary, preferably on large sized single crystalline samples.

*3.3. Hall effect*

The insert of Figure 5 shows the transverse resistivity $\rho_{xy}$ at different temperatures, which follows a linear relationship with the applied field and have a negative slope, $d\rho_{xy}/dH$. Also, the values of $\rho_{xy}$ are negative above $T_c$, indicating that the electric transport is dominated by electron-type carriers. From these data, $R_H = \rho_{xy}/H$ is determined and shown in Figure 5. The magnitude of $R_H$ decays continuously with the increasing temperature similar to that reported in single crystal NdFeAsO$_{0.82}$F$_{0.18}$ [32]. The strong temperature dependence of $R_H$ in iron-based superconductor is often simply attributed to the multiband effect because its Fermi surfaces contain two electron pockets and three hole pockets [18]. This hypothesis is reasonable for the undoped "1111" system or the "122" system, which contain both the electron and hole pockets. However, for the highly doped "1111" system, band structure calculations [33] show that with more than 10% electron doping, as is present in our sample, the hole pockets will shrink into a small ones and the electron pockets expand a lot. In this case the hole pockets have little effect on the electric conduction, thus the Hall coefficient can be expressed by a two electron band model:

$$eR_H = \frac{1}{n_H} = \frac{n_1 \mu_1^2 + n_2 \mu_2^2}{(n_1 \mu_1 + n_2 \mu_2)^2} = \frac{n_1 + n_2 (\mu_2/\mu_1)^2}{(n_1 + n_2 \mu_2/\mu_1)^2}$$

, where $n_1$, $n_2$ and $\mu_1$, $\mu_2$ are the charge-carrier density and the mobility of the two electron bands, respectively. The charge-carrier density depends upon the band structure, and the mobility is determined by both the effective mass $m^*$, which is determined by the band structure, and the scattering rate. Although the scattering rate is temperature dependent, it has little impact on the Hall coefficient, for its effect on both of the two electron bands is similar and thus can be excluded from $\mu_2/\mu_1$, similar to the one band model where the Hall coefficient is only dependent on the charge-carrier density. Thus the strong temperature dependent Hall coefficient can be attributed to a change of the band structure, which affects both the charge-carrier density and the effective mass. Furthermore, the change of the band structure may be attributed to a change on the Fermi surface supported by the pseudogap which causes the partial gapping of the Fermi surface. The pseudogap has actually been found from NMR [34] and photoemission experiments [35]; recently the quasiparticle relaxation dynamics on single crystal SmFeAsO$_{1-x}$F$_x$ [36] reveals the pseudogap with an onset above 180 K. The pseudogap, which is common in the cuprate superconductors, also causes the temperature dependent Hall coefficient associated with the depletion of the density of states (DOS).

To investigate the charge-carrier density, the two band model is simplified to a single band model assuming the two electron pockets are highly degenerated, which is reasonable from the calculated Fermi surface [33]. Thus the Hall coefficient can be expressed by $R_H = 1/ne$, where $n$ is the charge-carrier density, which is shown in Figure 6. The low charge-carrier density is also indicated by other experimental data [25] [32] [37] [38] and the electronic structure calculations [18]. From the measured Hall coefficient together with the London penetration depth $\lambda_0^2 = m^*/\mu_0 n e^2$ at $T=0$, the effective mass of the carriers can be roughly estimated as $m^* = \mu_0 e \lambda_0^2 / R_H$ [12], where $e$ is the electron charge. Taking $\lambda_0 = 189 nm$ from the μSR data [39], the results for the effective mass are calculated and also shown in Figure 6. The larger electronic effective mass reflects the renormalization caused by strong-coupling effects [33]. Considering the effective mass anisotropy parameter $\Gamma = m_c/m_{ab} = 18.6$ obtained above from the $H_{c2}$, a larger $m_c^*$ as well as a much smaller $m_{ab}^*$ can be deduced. Irrespective of the scattering rate, results of the effective mass show that the electronic mobility in the ab-plane is much larger than that of the c-axis.

## 4. Conclusion

In summary, we have prepared a high-quality sample of superconducting SmFeAsO$_{0.75}$, and systematically studied its electronic behavior by transport and specific heat measurements. An obvious specific heat jump associated with the superconducting transition was witnessed, and a very small normalized superconducting jump with the value of $\Delta C / \gamma_n T_c \sim 0.2$ was found and may be the result of various contributing factors. The temperature dependent Hall coefficient and the carrier density were obtained, which were attributed to the partial gapping of the Fermi surface caused by the emergence of the pseudogap.

## 5. Acknowledgment

We are very grateful to Dr. M. Sumption and Mr. Susner for discussions. This work was supported by the Natural Science Foundation of China, the Ministry of Science and Technology of China (973 project: No. 2011CBA00100), by the National Science Foundation of Jiangsu Province of China (Grant No. BK2010421) and by the Natural Science Foundation of China (No. 10904013).

# 5. Reference

[1] Y. Kamihara, T. Watanabe, M. Hirano, and H. Hosono, Journal of the American Chemical Society **130**, 3296 (2008).

[2] M. Rotter, M. Tegel, and D. Johrendt, Physical Review Letters **101,** 107006 (2008).

[3] X. Wang, Q. Liu, Y. Lv, W. Gao, L. Yang, R. Yu, F. Li, and C. Jin, Solid State Communications **148**, 538 (2008).

[4] F. Hsu, J. Luo, K. Yeh, T. Chen, T. Huang, P. Wu, Y. Lee, Y. Huang, Y. Chu, and D. Yan, Proceedings of the National Academy of Sciences **105**, 14262 (2008).

[5] Y. J. Jo, J. Jaroszynski, A. Yamamoto, A. Gurevich, S. C. Riggs, G. S. Boebinger, D. Larbalestier, H. H. Wen, N. D. Zhigadlo, S. Katrych, Z. Bukowski, J. Karpinski, R. H. Liu, H. Chen, X. H. Chen, and L. Balicas, Physica C: Superconductivity **469**, 566 (2009).

[6] Y. L. Wang, L. Shan, L. Fang, P. Cheng, C. Ren and Hai-Hu Wen, Superconductor Science and Technology **22**, 015018 (2009).

[7] C. Ren, Z. S. Wang, H. Q. Luo, H. Yang, L. Shan, and H. H. Wen, Physica C: Superconductivity **469**, 599 (2009).

[8] O. Millo, I. Asulin, O. Yuli, I. Felner, Z. A. Ren, X. L. Shen, G. C. Che, and Z. X. Zhao, Physical Review B **78**, 092505 (2008).

[9] L. Ding, C. He, J. K. Dong, T. Wu, R. H. Liu, X. H. Chen, and S. Y. Li, Physical Review B **77**, 180510 (2008).

[10] M. Tropeano, A. Martinelli, A. Palenzona, E. Bellingeri, E. Galleani d'Agliano, T. D. Nguyen, M. Affronte, and M. Putti, Physical Review B **78**, 094518 (2008).

[11] R. H. Liu, G. Wu, T. Wu, D. F. Fang, H. Chen, S. Y. Li, K. Liu, Y. L. Xie, X. F. Wang, R. L. Yang, L. Ding, C. He, D. L. Feng, and X. H. Chen, Physical Review Letters **101**, 087001 (2008).

[12] J. Jaroszynski, S. C. Riggs, F. Hunte, A. Gurevich, D. C. Larbalestier, G. S. Boebinger, F. F. Balakirev, A. Migliori, Z. A. Ren, W. Lu, J. Yang, X. L. Shen, X. L. Dong, Z. X. Zhao, R. Jin, A. S. Sefat, M. A. McGuire, B. C. Sales, D. K. Christen, and D. Mandrus, Physical Review B **78**, 064511 (2008).

[13] R. Zhi-An and et al., Chinese Physics Letters **25**, 2215 (2008).

[14] H. Lee, J. Park, J. Lee, J. Kim, N. Sung, T. Koo, B. Cho, C. Jung, S. Saini, and S. Kim, Superconductor Science and Technology **22**, 075023 (2009).

[15] ouml, H. v. hneysen, A. Rosch, M. Vojta, and P. lfle, Reviews of Modern Physics **79**, 1015 (2007).

[16] H.-S. Lee, M. Bartkowiak, J. H. Park, J. Y. Lee, J. Y. Kim, N.-H. Sung, B. K. Cho, C.-U. Jung, J. S. Kim, and H. J. Lee, Physical Review B **80**, 144512 (2009).

[17] P. C. Y. Jia, L. Fang, H. Q. Luo, H. Yang, C. Ren, L. Shan, C. Z. Gu, and H. H. Wen, Appl. Phys. Lett. **93**, 032503 (2008).

[18] D. J. Singh and M. H. Du, Physical Review Letters **100**, 237003 (2008).

[19] K. K. Nanda, Physica C: Superconductivity **265**, 26 (1996).

[20] G. Mu, X. Y. Zhu, L. Fang, L. Shan, C. Ren, and H. H. Wen, Chinese Physics Letters **25**, 2221 (2008).

[21] P. J. Baker, S. R. Giblin, F. L. Pratt, R. H. Liu, G. Wu, X. H. Chen, M. J. Pitcher, D. R. Parker, S. J. Clarke, and S. J. Blundell, New Journal of Physics **11**, 025010 (2009).

[22] S. Riggs, C. Tarantini, J. Jaroszynski, A. Gurevich, A. Palenzona, M. Putti, T. D. Nguyen, and M. Affronte, Physical Review B **80**, 214404 (2009).

[23] V. Tsurkan, J. Deisenhofer, A. Günther, Ch. Kant, H.-A. Krug von Nidda, F. Schrettle, A. Loidl, arXiv:1006.4453v2 (2010).

[24] K. Yoshimitsu, K. Yoichi, K. Hitoshi, A. Tooru, H. Masahiro, and H. Hideo, Journal of the Physical Society of Japan **77**, 094715 (2008).

[25] S. Suzuki, S. Miyasaka, S. Tajima, T. Kida, and M. Hagiwara, Journal of the Physical Society of Japan **78**, 114712 (2009).

[26] G. Mu, H. Q. Luo, Z. S. Wang, L. Shan, C. Ren, and H. H. Wen, Physical Review B **79**, 174501 (2009).

[27] Z. Li, G. Chen, J. Dong, G. Li, W. Hu, D. Wu, S. Su, P. Zheng, T. Xiang, N. Wang, and J. Luo, Physical Review B **78**, 060504 (2008).

[28] J. T. Park, D. S. Inosov, C. Niedermayer, G. L. Sun, D. Haug, N. B. Christensen, R. Dinnebier, A. V. Boris, A. J. Drew, L. Schulz, T. Shapoval, U. Wolff, V. Neu, X. Yang, C. T. Lin, B. Keimer, and V. Hinkov, Physical Review Letters **102**, 117006 (2009).

[29] H. Mukuda, N. Terasaki, N. Tamura, H. Kinouchi, M. Yashima, Y. Kitaoka, K. Miyazawa, P. M. Shirage, S. Suzuki, S. Miyasaka, S. Tajima, H. Kito, H. Eisaki, and A. Iyo, Journal of the Physical Society of Japan **78** (2009).

[30] J. Yang, Z. A. Ren, G. C. Che, W. Lu, X. L. Shen, Z. C. Li, W. Yi, X. L. Dong, L. L. Sun, F. Zhou, and Z. X. Zhao, Superconductor Science & Technology **22**, 025004 (2009).

[31] H. H. Wen, G. Mu, H. Luo, H. Yang, L. Shan, C. Ren, P. Cheng, J. Yan, and L. Fang, Physical Review Letters **103**, 067002 (2009).

[32] P. Cheng, H. Yang, Y. Jia, L. Fang, X. Y. Zhu, G. Mu, and H. H. Wen, Physical Review B **78** 134508 (2008).

[33] I. I. Mazin, D. J. Singh, M. D. Johannes, and M. H. Du, Physical Review Letters **101**, 057003 (2008).

[34] H. J. Grafe, D. Paar, G. Lang, N. J. Curro, G. Behr, J. Werner, J. Hamann-Borrero, C. Hess, N. Leps, R. Klingeler, uuml, and B. chner, Physical Review Letters **101**, 047003 (2008).

[35] L. Hai-Yun and et al., Chinese Physics Letters **25**, 3761 (2008).

[36] T. Mertelj, P. Kusar, V. V. Kabanov, L. Stojchevska, N. D. Zhigadlo, S. Katrych, Z. Bukowski, J. Karpinski, S. Weyeneth, and D. Mihailovic, Physical Review B **81**, 224504 (2010).

[37] G. F. Chen, Z. Li, G. Li, J. Zhou, D. Wu, J. Dong, W. Z. Hu, P. Zheng, Z. J. Chen, H. Q. Yuan, J. Singleton, J. L. Luo, and N. L. Wang, Physical Review Letters **101**, 057007 (2008).

[38] Xiyu Zhu , Lei Fang , Gang Mu and Hai-Hu Wen Supercond. Sci. Technol **21**, 105001 (2008).

[39] A. J. Drew, F. L. Pratt, T. Lancaster, S. J. Blundell, P. J. Baker, R. H. Liu, G. Wu, X. H. Chen, I. Watanabe, V. K. Malik, A. Dubroka, K. W. Kim, ouml, M. ssle, and C. Bernhard, Physical Review Letters **101**, 097010 (2008).

# Figure captions

Figure 1: Temperature dependence of the electrical resistivity for SmFeAsO$_{0.75}$. The solid line is the linear fitting of the resistivity. The inset is the resistivity broadening of SmFeAsO$_{0.75}$ with increasing magnetic field up to 9 T.

Figure 2: Temperature dependence of specific heat at zero field. The solid line is the fitting curve of the normal state specific heat by combined Einstein-Debye model. The lower inset shows the enlarged part of superconducting transition of the specific heat plotted as *C/T* vs *T*. The upper inset is the enlarged low temperature specific heat peak at the field of 0 T and 9 T.

Figure 3: The temperature dependence of specific heat at low temperatures for 0 T and 9 T plotted as *C/T* vs $T^2$. The solid lines are the linear fits between 14 and 28 K.

Figure 4: Temperature dependence of the specific heat close to $T_c$ in 0 T and 9 T plotted as *C/T* vs *T*. The inset shows $[C(0T)-C(9T)]/T$ vs temperature.

Figure 5: Temperature dependence of the Hall coefficient $R_H$. The inset shows a good linear relationship between $\rho_{xy}$ and *H* at different temperatures.

Figure 6: Temperature dependence of the charge-carrier density $n$ and the effective mass $m^*$.

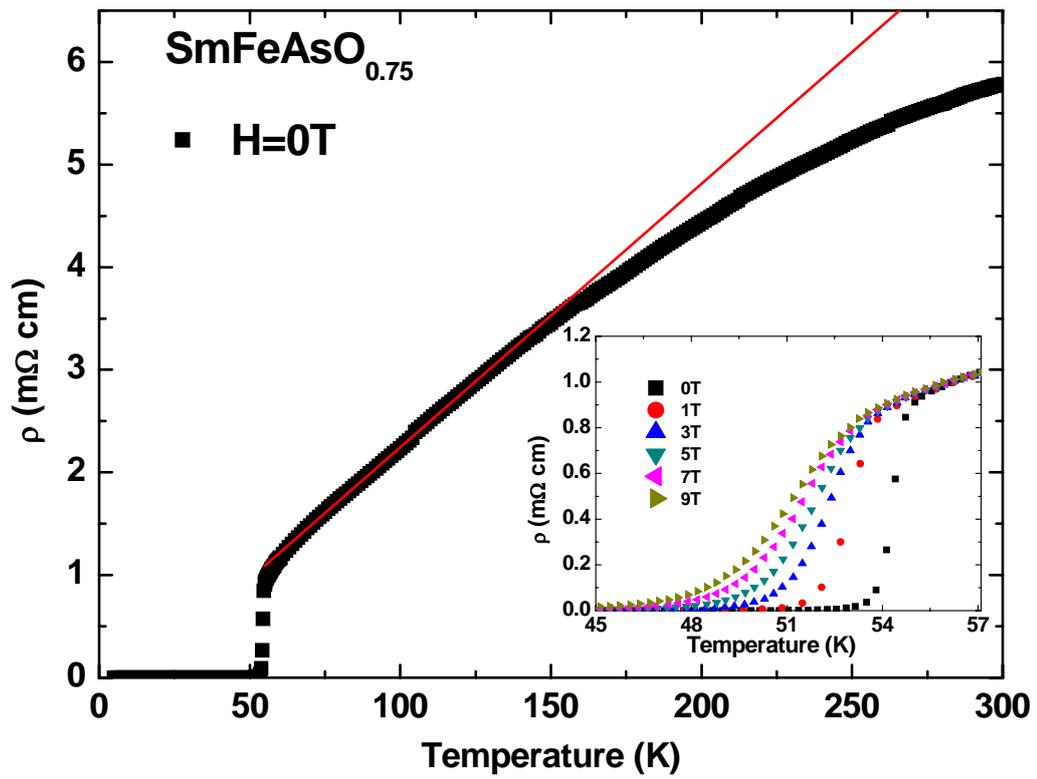

**Figure. 1** Y. Sun *et al.*

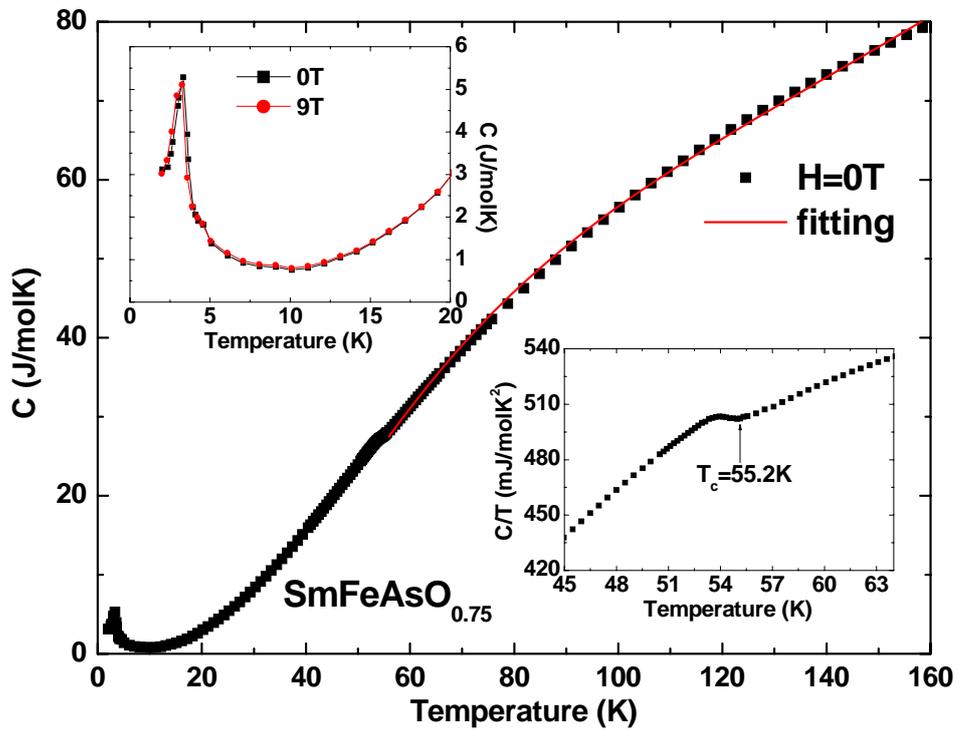

**Figure. 2 Y. Sun** *et al.*

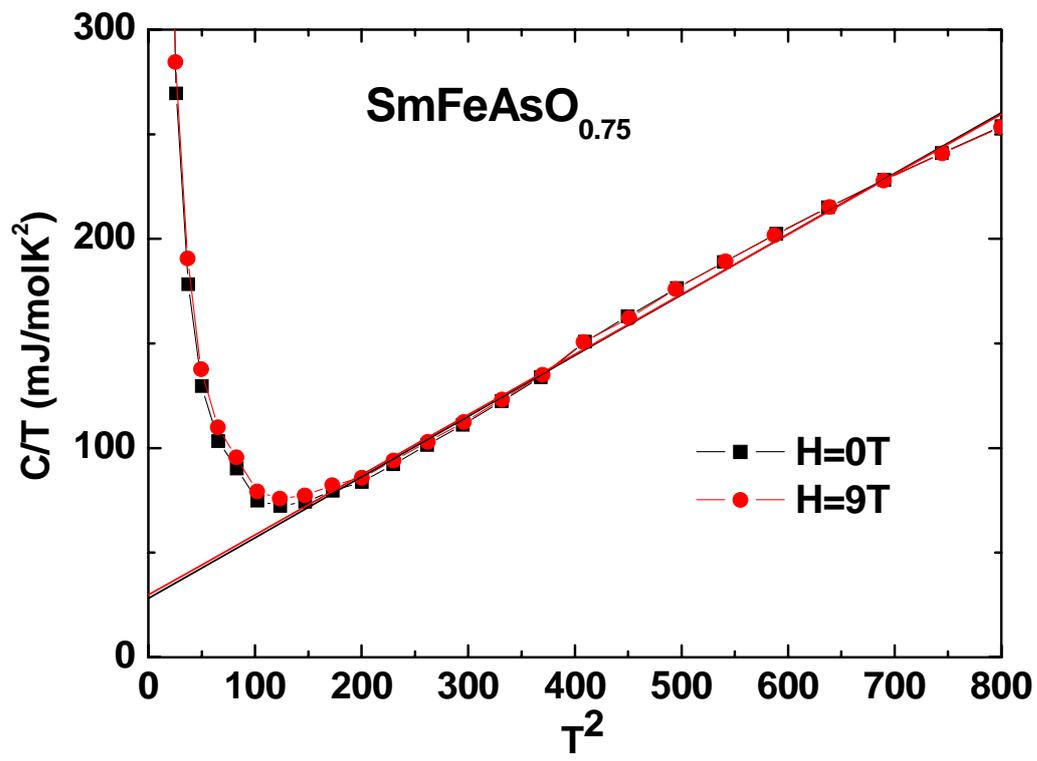

Figure. 3 Y. Sun *et al.*

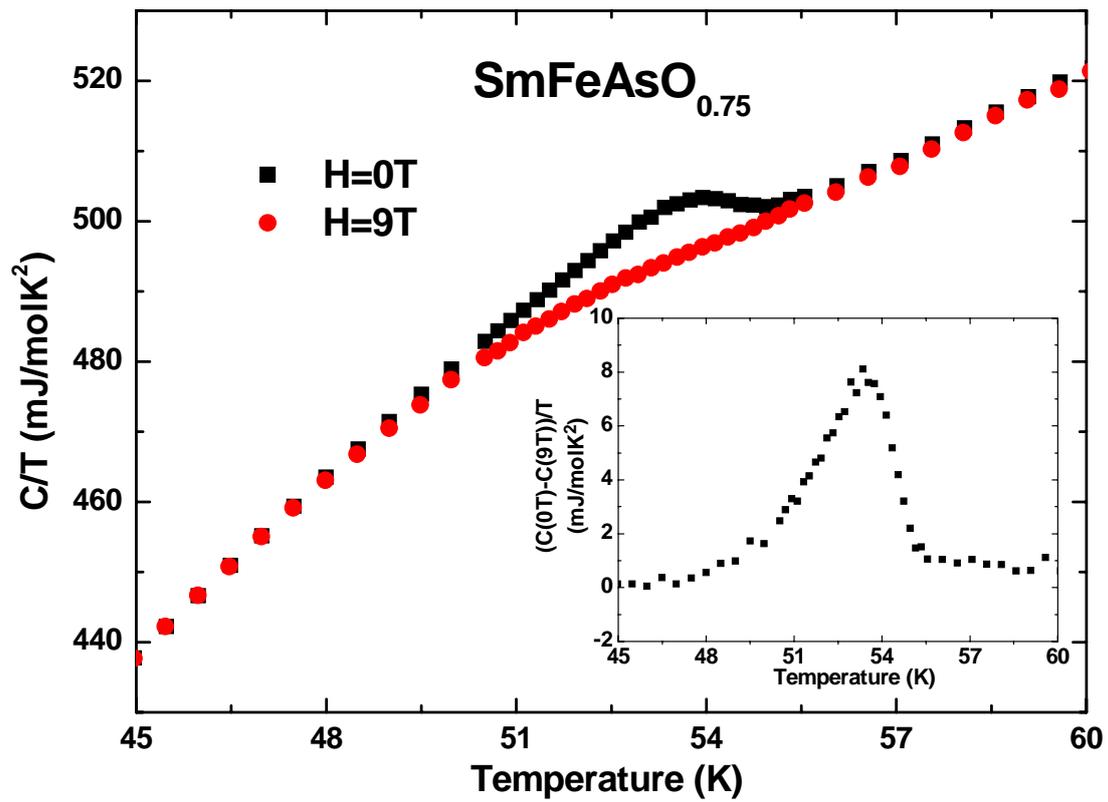

**Figure. 4** Y. Sun *et al.*

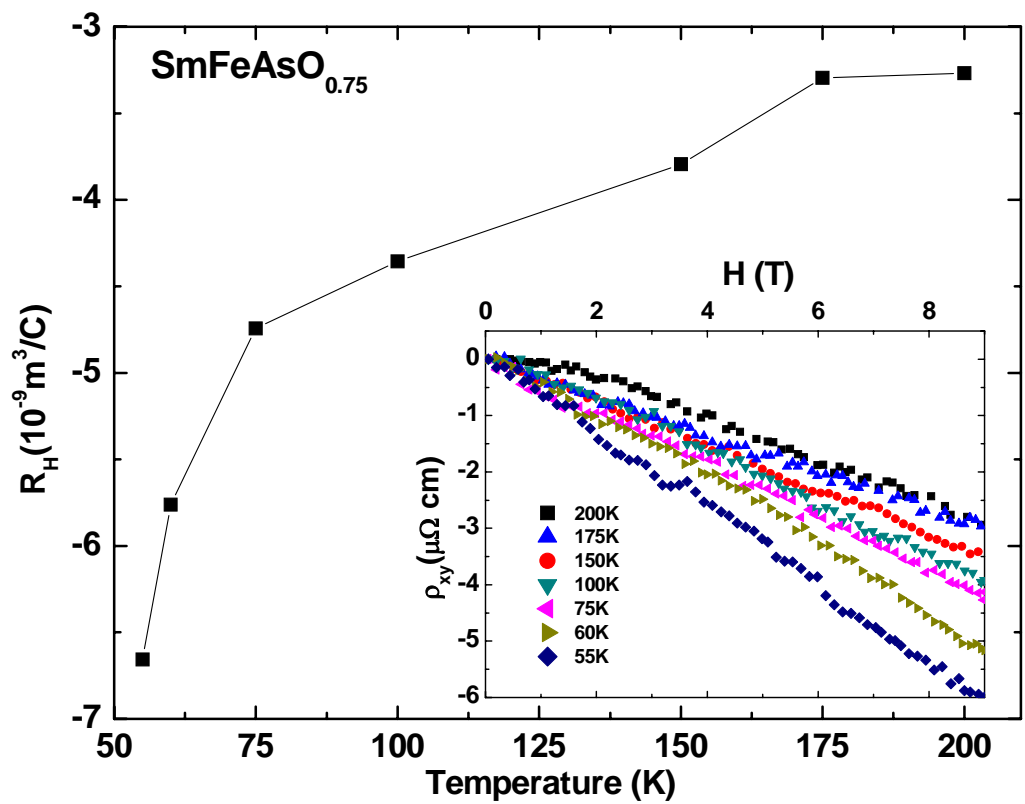

**Figure. 5** Y. Sun *et al.*

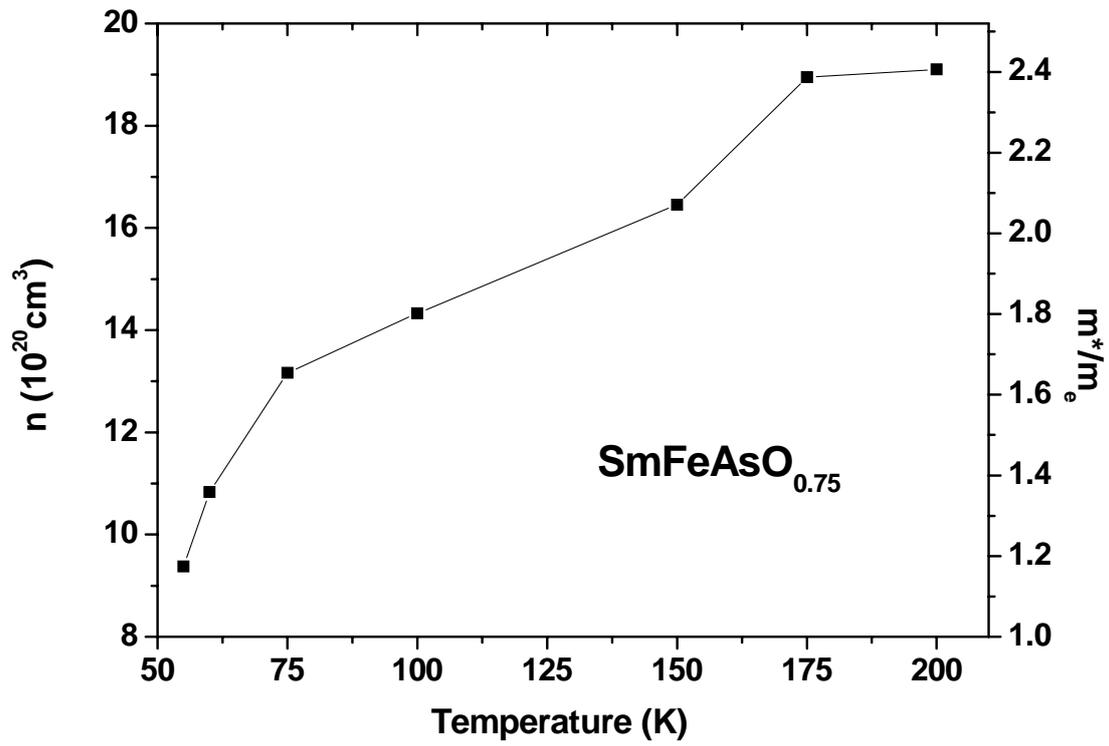

**Figure. 6** Y. Sun *et al.*